\documentclass{iopart}
\usepackage[dvips]{graphicx}

\def\ba{\begin{array}}
\def\ea{\end{array}}

\begin{document}
\title[Matrix Product States of Interacting
Particle Systems] {Matrix Product States of Three Families of
One-Dimensional Interacting Particle Systems}
\author{Farhad H Jafarpour \footnote[1]{e-mail:farhad@sadaf.basu.ac.ir}}
\address{Bu-Ali Sina University, Physics Department, Hamadan, Iran \\
Institute for Studies in Theoretical Physics and Mathematics
(IPM),\\ P.O. Box 19395-5531, Tehran, Iran}

\begin{abstract}
The steady states of three families of one-dimensional
non-equilibrium models with open boundaries, first proposed in
\cite{kjs}, are studied using a matrix product formalism. It is
shown that their associated quadratic algebras have
two-dimensional representations, provided that the transition
rates lie on specific manifolds of parameters . Exact expressions
for the correlation functions of each model have also been
obtained. We have also studied the steady state properties of one
of these models, first introduced in \cite{farhad2}, with
more details. By introducing a canonical ensemble we calculate the
canonical partition function of this model exactly. Using the
Yang-Lee theory of phase transitions we spot a second-order phase
transition from a power-law to a jammed phase. The density profile
of particles in each phase has also been studied. A simple
generalization of this model in which both the left and the right
boundaries are open has also been introduced. It is shown that
double shock structures may evolve in the system under certain
conditions.
\end{abstract}
\pacs{02.50.Ey, 02.50.Ga, 05.20.-y, 05.70.Fh, 05.70.Ln}
\maketitle

\section{Introduction}

The occurrence of non-equilibrium phase transitions is one of the
most interesting properties of one-dimensional out-of-equilibrium
models \cite{sz,sch}. In open boundary systems, changing the
boundary parameters may result in special kinds of non-equilibrium
phase transitions which are called {\it the boundary induced phase
transitions}. The existence of shocks is another interesting
property of these lattice models. A shock is defined as a
discontinuity between a hight-density and a low-density region in
the density profile of particles on the lattice. The shocks have
been studied both at the macroscopic level using the hydrodynamic
equations and also at the microscopic level by solving the master
equation of spatially discrete models. For example, at the
macroscopic scale the shocks may appear in the solutions of the
Burgers equation with particles conversation \cite{burgers}. At
the microscopic level the Burgers equation can be described by the
Asymmetric Simple Exclusion Process (ASEP) \cite{liggett}. The
ASEP cotains one class of particles which can be injected and
extracted from the boundaries of a one-dimensional chain while
hopping in the bulk with asymmetric rates. This model has been
studied widely during the last decade. It is known that by
changing the injection and extraction rates, the ASEP undergoes
several boundary induced phase transitions \cite{dehp}. The shocks
may also appear in the density profile of the particles with the
ASEP dynamics for certain values of the reaction rates. The same
phenomenon might happen when the fixed number of such particles
hop on a ring in the presence of a slow particle
called an impurity \cite{DJLS}-\cite{farhad1}.\\
Recently, it has been shown that among one-dimensional
reaction-diffusion systems with open boundaries there are three
families of models with a common property: The time evolution
equation of a Bernoulli shock measure developed by the Hamiltonian
of each of these models is quite similar to the time evolution
equation of a single-particle random walk on a lattice with
reflecting boundaries \cite{kjs}. This actually takes place
provided that some constraints on the reaction rates of these
systems are fulfilled. These models are the Asymmetric Simple
Exclusion Process (ASEP), the Branching-Coalescing Random Walk
(BCRW) and the Asymmetric Kawasaki-Glauber process (AKGP). It is
known that in an infinite system, and under some constraints,
single shock or even consecutive multiple shocks may evolve in the
ASEP \cite{ferrari1}-\cite{ferrari2}. In the latter case the time
evolution equation for the multiple shock measure generated by the
Hamiltonian of the system is similar to the time evolution
equation of $n$ random walkers. In the steady state the ASEP's
probability distribution function can be written in terms of
superposition of shocks. From the Matrix Product Formalism (MPF)
point of view, in which the steady state weights are written in
terms of the expectation values of product of non-commuting
operators \cite{dehp}, $n$ consecutive shocks correspond to an
$(n+1)$-dimensional representation of the ASEP's quadratic
algebra \cite{essrit,malsan}.\\
The question that might arise is that whether or not the steady
state of the BCRW and that of the AKGP can also be represented by
finite dimensional representation of their quadratic algebras
since their steady states can be written in terms of interactions
of single shock. We will deal with this question in the first part
of this paper. We will show that as far as a single shock
structure is concerned the steady state of the BCRW and that of
the AKGP can be obtained from two-dimensional representations
of their quadratic algebras.\\
In \cite{farhad2} we have studied a branching-coalescing model
with open boundary in which the particles diffuse, coagulate and
decoagulate on a one-dimensional chain. The particles are also
allowed to enter and leave the chain only from the left boundary.
This model had already been studied with reflecting boundaries in
\cite{hkp1,hkp2}. We have shown that the steady state of this open
boundary model can be obtained using the MPF provided that the
reaction rates satisfy a constraint. In this case its quadratic
algebra has a two-dimensional representation, while in the case of
reflecting boundary conditions the quadratic algebra of the model
has a four-dimensional representation. Recent investigations show
that some of the classical concepts and theories in the
equilibrium statistical mechanics, such as the Yang-Lee theory of
equilibrium phase transitions \cite{yanglee}, can be successfully
applied to the out-of-equilibrium systems. There are several
examples from the driven diffusive systems to the directed
percolation models which confirm the applicability of this theory
to the non-equilibrium systems \cite{farhad2}-\cite{bcdl}. We have
also studied the phase transition of our model using the Yang-Lee
theory by studying the roots of its grand canonical partition
function \cite{farhad2}. These zeros lie on a circle and intersect
the positive real axis at an angle $\frac{\pi}{2}$. According to
the Yang-Lee theory this reveals a first-order phase transition
\cite{gross}. As we will see this model is an special case of the
BCRW; therefore, we expect to see the shock profiles in its steady
state. In order to see the shock profiles we will confine the
number of particles by working in a canonical ensemble. The sum
over the steady state weights with fixed number of particles is
defined as the canonical partition function of this model. The
canonical partition function can be calculated exactly using the
MPF. It turns out that the zeros of this function in the complex
plan of one of the parameters lie on a curve which intersects the
positive real axis at an angle $\frac{\pi}{4}$. At the same time
their density vanishes near the critical point. It indicates a
second-order phase transition. In the canonical ensemble one can
also calculate the density profile of particles on the chain again
using the MPF. Our calculations show that in one phase the density
profile of particle has an exponential behavior while in the other
phase, as we expect, it is a shock.\\ Another question that might
arise is that under what conditions multiple shocks may evolve in
this system. We will show that in our model if we let the
particles to enter and leave the chain from the both ends of the
chain, one can see the double shock structures given that the
injection and extraction rates satisfy some constraints. In this
case the time evolution equation of the shock measure is similar
to that of two random walkers. This phenomenon has already been
observed in ASEP \cite{kjs,beli,pigo}.\\
This paper is organized as follows: In the second section we will
briefly review the basic concepts of the MPF. In the third section
we will show that the quadratic algebra of the BCRW, as a
generalized form of our branching-coalescing model, has a
two-dimensional representation provided that the same constraints
introduced in \cite{kjs} are fulfilled. In this section the
two-dimensional representation of the AKGP will also be
introduced. In the forth section we will switch to our model
proposed in \cite{farhad2} and find its canonical partition
function using the MPF. The numerical estimates of the canonical
partition function zeros, as a function of a complex reaction
rate, will be obtained. The line of zeros and also their density
will be calculated exactly. Finally, we will calculate the density
profile of particles in each phase. The fifth section is devoted
to the study of double shock structures in a general
branching-coalescing model with open boundaries. In the last
section we will summarize our results.

\section{The matrix product formalism}

In this section we review the basic ideas of the Matrix Product
Formalism (MPF) first introduced in \cite{dehp}. We will consider
only one-dimensional stochastic systems with nearest neighbors
interactions and open boundaries. Systems with long range
interactions and open
boundaries have also been studied in related literatures \cite{ks}.\\
Let us define $P({\cal C};t)$ as the probability distribution of
any configuration $\cal C$ of a Markovian interacting particle
system at the time $t$. The time evolution of $P({\cal C};t)$ can
be written as a Schr\"odinger equation in imaginary time
\begin{equation}
\frac{d}{dt}P({\cal C};t)=H P({\cal C};t)
\end{equation}
in which $H$ is a stochastic Hamiltonian whose matrix elements are
the transition rates between different configurations. For the
one-dimensional systems of length $L$ which consist of one species
of particles the Hamiltonian $H$ has the following general form
\begin{equation}
H=\sum_{k=1}^{L-1}h_{k,k+1}+h_1+h_L
\end{equation}
in which
$$
h_{k,k+1}={\cal I}^{\otimes (k-1)}\otimes h \otimes {\cal
I}^{\otimes (L-k-1)} \;,\; h_1=h^{(l)} \otimes {\cal I}^{\otimes
(L-1)} \;,\; h_L={\cal I}^{\otimes (L-1)}\otimes h^{(r)}
$$
where ${\cal I}$ is a $2 \times 2$ identity matrix, $h$ is a $4
\times 4$ matrix for the bulk interactions and $h^{(l)}$
($h^{(r)}$) is a $2 \times 2$ matrix for particle input and output
from the left (right) boundary. In a basis $(00,01,10,11)$ the
bulk Hamiltonian along with the boundary Hamiltonians are given by
\begin{equation}
\begin{array}{c}
\label{h1}
h = \left( \begin{array}{cccc}
.& w_{12} & w_{13} & w_{14} \\
w_{21} & . & w_{23} & w_{24} \\
w_{31} & w_{32} & . & w_{34} \\
w_{41} & w_{42} & w_{43} & . \end{array} \right),\\
 h^{(l)} = \left(
\begin{array}{rr}
-\alpha & \gamma \\
\alpha & -\gamma \end{array} \right),h^{(r)} =  \left(
\begin{array}{rr}
-\delta & \beta \\
\delta & -\beta \end{array} \right).
\end{array}
\end{equation}
Requiring the conservation of probability, the diagonal terms of
matrices should be the negative of the sum of transition rates in
each column. In the most general form the interaction rates are
\begin{equation}
\label{r1}
\begin{array}{ll}
\mbox{Diffusion to the left and right } & w_{32}, \; w_{23} \\
\mbox{Coalescence to the left and right } & w_{34}, \; w_{24} \\
\mbox{Branching to the left and right } & w_{43}, \; w_{42} \\
\mbox{Death to the left and right }  & w_{13}, \; w_{12} \\
\mbox{Birth to the left and right } & w_{31}, \; w_{21}  \\
\mbox{Pair Annihilation and Creation }& w_{14}, \; w_{41}\\
\mbox{Injection and Extraction at the first site} & \alpha, \; \gamma  \\
\mbox{Injection and Extraction at the last site}  & \delta, \;
\beta.
\end{array}
\end{equation}
In the stationary state we have $H P^*({\cal C})=0$. According to
the MPF the stationary probability distribution $P^*({\cal C})$ is
assumed to be of the form
\begin{equation}
\label{Weigth} P^*({\cal
C})=P^*({\tau_1,\cdots,\tau_L})=\frac{1}{Z_{L}} \langle W \vert
\prod_{i=1}^{L}(\tau_i D+(1-\tau_i)E)\vert V \rangle
\end{equation}
where $\tau_i=0$ if the site $i$ is empty and $\tau_i=1$ if it is
occupied by a particle. The function $Z_L$ in (\ref{Weigth}) is a
normalization factor and can be obtained easily using the
normalization condition $\sum_{\cal C}P^*({\cal C})=1$. The
operators $D$ and $E$ stand for the presence of a particle and a
vacancy at each site. These operators besides the vectors $\langle
W \vert$ and $\vert V \rangle$ should satisfy the following
algebra
\begin{equation}
\begin{array}{l}
\label{MPA}
h \left[ \left( \begin{array}{c} E \\
D \end{array} \right) \otimes
\left( \begin{array}{c} E \\
D \end{array} \right) \right]=
\left( \begin{array}{c} \bar{E} \\
 \bar{D} \end{array} \right) \otimes
\left( \begin{array}{c} E \\
 D \end{array} \right) -
\left( \begin{array}{c} E \\
 D \end{array} \right) \otimes
\left( \begin{array}{c} \bar{E} \\
 \bar{D} \end{array} \right), \\ \\
\langle W| \; h^{(l)} \left( \begin{array}{c} E \\
 D \end{array} \right) =
-\langle W| \left( \begin{array}{c} \bar{E} \\
 \bar{D} \end{array} \right),\\ \\
h^{(r)} \left( \begin{array}{c} E \\
 D \end{array} \right) |V \rangle =
\left(\begin{array}{c} \bar{E} \\
 \bar{D} \end{array} \right) |V \rangle. \nonumber
\end{array}
\end{equation}
The operators $\bar{E}$ and $\bar{D}$ are auxiliary operators and
do not enter in the calculation of physical quantities. Using
(\ref{h1}) and (\ref{MPA}) the quadratic algebra associated with
the most general reaction-diffusion model can be obtained. By
defining $C:=D+E$ one can easily see that the mean values of the
physical quantities can be written in terms of the non-commuting
operators $C$ and $E$ and the vectors $\vert V \rangle$ and
$\langle W \vert$. For example the mean density of particles at
site $i$ has the following form:
\begin{eqnarray}
\label{OPF} \langle \rho_i \rangle=\frac{\sum_{\cal C}\tau_i
P^*({\cal C})}{\sum_{\cal C}P^*({\cal C})} =\frac{\langle W \vert
C^{i-1} (C-E) C^{L-i}\vert V\rangle}{\langle C^L \rangle}.
\end{eqnarray}
Similarly, any $n$-point density correlation function can be
written as
\begin{eqnarray}
\label{NPF}
\langle\rho_{i_1}\rho_{i_2}\cdots\rho_{i_n}\rangle=\frac{\sum_{\cal
C}\tau_{i_1} \tau_{i_2}\cdots \tau_{i_n} P^*({\cal C})}{\sum_{\cal
C}P^*({\cal C})} \nonumber \\
=\frac{\langle W \vert C^{i_1-1}(C-E)C^{i_2-i_1-1}\cdots
(C-E)C^{L-i_n}\vert V\rangle}{\langle C^L \rangle}.
\end{eqnarray}
In order to calculate (\ref{OPF}), (\ref{NPF}), or any other
quantity, there are two possibilities: one might find a finite or
an infinite-dimensional representation for the quadratic algebra
of the system or equivalently, one may use the commutation
relation between the operators to calculate these quantities
rigorously . In the next section we will show that
finite-dimensional representation exists for three families of
reaction-diffusion systems, given that some constraints are
satisfied.

\section{Representation of the quadratic algebra}

In our recent paper \cite{kjs} exact travelling wave solutions are
obtained for three families of one-dimensional reaction-diffusion
models with open boundaries. These models are the Asymmetric
Simple Exclusion Process (ASEP), the Branching-Coalescing Random
Walk (BCRW) and the Asymmetric Kawasaki-Glauber Process (AKGP). It
has been shown that for these models the stationary measure can be
written as a linear combination of Bernoulli shock measures
defined on a lattice of length $L$ as
\begin{equation}
\label{BSM}
\vert m \rangle= \left( \ba{c} 1-\rho_1 \\
\rho_1 \ea \right)^{\otimes m} \otimes \left( \ba{c} 1-\rho_2 \\
\rho_2 \ea \right)^{\otimes L-m} \; \; , \; \; 0\leq m \leq L
\end{equation}
provided that some constraints on the reaction rates are
fulfilled. In (\ref{BSM}) $\rho_1$ and $\rho_2$ are the densities
of particles at the left and the right domains of the shock
position $m$ respectively. The time evolution of (\ref{BSM})
generated by the Hamiltonian of the above-mentioned models is
given by
\begin{equation}
- H \vert m \rangle = d_1 \vert m-1 \rangle + d_2 \vert m+1
\rangle - (d_1+d_2) \vert m \rangle \; \; , \; \; 0 < m < L
\end{equation}
which is a simple single-particle random walk equation for the
position of the shock $m$ with hopping rate $d_1$ to the left and
$d_2$ to the right \cite{kjs}. In this section we show that under
the same constraints the stationary states of these models can be
studied using the MPF and their associated quadratic algebras have
two-dimensional representations.\\
For the ASEP the non-vanishing rates in (\ref{r1}) are $w_{32}$,
$w_{23}$, $\alpha$, $\beta$, $\gamma$ and $\delta$. It is known
that the quadratic algebra of the ASEP has an $n$-dimensional
representation $(n \geq 2)$, provided that the parameters of the
model satisfy some constraints \cite{essrit,malsan}. The
$(n+1)$-dimensional representations of the quadratic algebra
describe the stationary linear combination of shock measures with
$n$ consecutive shocks. The ASEP has already been studied widely
in literatures; therefore, will not be considered here
(see \cite{sch} and references therein).\\
For the BCRW the non-vanishing parameters in (\ref{r1}) are
$w_{34}$, $w_{24}$, $w_{42}$, $w_{43}$, $w_{32}$, $w_{23}$,
$\alpha$, $\beta$ and $\gamma$. As we will see, the steady state
of this model can be described by two-dimensional representation
of its quadratic algebra given that
\begin{equation}
\label{Rules}
\begin{array}{l}
\frac{1-\rho}{\rho}=\frac{w_{24}+w_{34}}{w_{42}+w_{43}}\;,\;
w_{23}= \frac{1-\rho}{\rho}w_{43}\\
\gamma= \frac{1-\rho}{\rho}\alpha + (1-\rho) w_{32}-
\frac{1-\rho}{\rho} w_{43} + \rho w_{34}
\end{array}
\end{equation}
in which $\rho_1=\rho$ is the density of particles on the
left-hand side of the shock. The density of particles on the
right-hand side of the shock $\rho_2$ is zero.\\
For the AKGP, the non-vanishing parameters in (\ref{r1}) are
$w_{12}$, $w_{13}$, $w_{42}$, $w_{43}$, $w_{32}$, $\alpha$ and
$\beta$; however, there is no specific constraint on these
parameters in order to have the mentioned property. If the
particles are allowed only to enter the system from the first site
with the rate $\alpha$ and leave it from the last site of the
chain with the rate $\beta$, then we will have $\rho_1=1$ and $\rho_2=0$.\\
We have found that the quadratic algebras of the BCRW and the AKGP
have two-dimensional representations given that the constraints
(\ref{Rules}) for the BCRW are fulfilled. For $c_{11}\neq c_{22}$
we find
\begin{equation}
\begin{array}{c}
\label{Rep1}
C=\left(\begin{array}{cc}
c_{11} & 0  \\
0 & c_{22}
\end{array} \right)\; , \;
E=\left(\begin{array}{cc}
e_{11} & e_{12}  \\
0 & e_{22}
\end{array} \right) \\ \\
\vert V \rangle=\left(\begin{array}{c} v_1\\v_2
\end{array} \right) \; , \; \langle W \vert=\Bigl( w_1,w_2 \Bigr)
\end{array}
\end{equation}
and for $\tilde{c}=c_{11}=c_{22}$ we find
\begin{equation}
\begin{array}{c}
\label{Rep2} C=\left(\begin{array}{cc}
\tilde{c} & \tilde{c}_{12}  \\
0 & \tilde{c}
\end{array} \right) \; , \;
E=\left(\begin{array}{cc}
\tilde{e}_{11} & \tilde{e}_{12}  \\
0 & \tilde{e}_{22}
\end{array} \right)\\ \\
\vert V \rangle=\left(\begin{array}{c} \tilde{v}_1\\
\tilde{v}_2
\end{array} \right) \; , \; \langle W
\vert=\Bigl(\tilde{w}_1,\tilde{w}_2\Bigr).
\end{array}
\end{equation}
The matrix elements are functions of the non-vanishing parameters
of each model. The explicate form of the matrices $C$, $E$, $\bar{C}$
and $\bar{E}$ and the vectors $\vert V \rangle$ and $\langle W \vert$
for each model is given in Appendix. Using the representation
(\ref{Rep1}) for $c_{11}\neq c_{22}$ and the definition of the
density profile of particles (\ref{OPF}) one finds in the
thermodynamic limit ($L\rightarrow \infty$)
\begin{equation}
\label{OPFF1}\langle \rho_i
\rangle=\cases{(1-\frac{e_{11}}{c_{11}})-\frac{v_2e_{12}}{v_1
c_{11}}{e}^{\frac{i-L}{\xi}},&$c_{11}
>c_{22}$ \cr
(1-\frac{e_{22}}{c_{22}})-\frac{w_1e_{12}}{w_2c_{11}}{e}^{\frac{-i}{\xi}}
,&$c_{11}<c_{22}.$\cr}
\end{equation}
in which we have defined the characteristic length $\xi=\vert
\ln \frac{c_{11}}{c_{22}}\vert^{-1}$. Also using (\ref{Rep2}) and
(\ref{OPF}) for $\tilde{c}=c_{11}=c_{22}$
\begin{equation}
\label{OPFF2} \langle \rho_i
\rangle=(1-\frac{\tilde{e}_{11}}{\tilde{c}})+(\frac{\tilde{e}_{11}}{\tilde{c}}-
\frac{\tilde{e}_{22}}{\tilde{c}})x
\; \; , \; \; 0\leq x\leq 1
\end{equation}
where $x=\frac{i}{L}$. As can be seen the density profile of
particles has an exponential behavior for $c_{11}\neq c_{22}$
while it is linear on the coexistence line $c_{11}=c_{22}$. For
$c_{11}>c_{22}$ the density of particles near the left boundary is
$\rho_1=1-\frac{e_{11}}{c_{11}}$ while for $c_{11}<c_{22}$ the
density of particles near the right
boundary is $\rho_2=1-\frac{e_{22}}{c_{22}}$.\\
Having the two-dimensional representations (\ref{Rep1}) and
(\ref{Rep2}) any higher order correlation functions can easily be
calculated using (\ref{NPF}). For instance, in the thermodynamic
limit the two-point correlation function is obtained to be
\begin{eqnarray}
\label{tpcf1} \langle\rho_{i}\rho_{j}\rangle=\cases{
(\frac{d_{11}}{c_{11}})^2+\frac{v_2d_{11}d_{12}}{v_1c_{11}^2}e^{\frac{j-L}{\xi}}+
\frac{v_2d_{12}d_{22}}{v_1c_{11}c_{22}}e^{\frac{i-L}{\xi}},c_{11}>c_{22}&\cr
(\frac{d_{22}}{c_{22}})^2+\frac{w_1d_{11}d_{12}}{w_2c_{11}^2}e^{\frac{-j}{\xi}}+
\frac{w_1d_{12}d_{22}}{w_2c_{11}c_{22}}e^{\frac{-j}{\xi}},c_{11}<c_{22}.&\cr}
\end{eqnarray}
for $c_{11}\neq c_{22}$ and
\begin{equation}
\label{tpcf2}
\langle\rho_{i}\rho_{j}\rangle=(\frac{\tilde{d}_{11}}{\tilde{c}})^2+\frac{\tilde{d}_{11}
(\tilde{d}_{22}-\tilde{d}_{11})}{\tilde{c}^2}
(\frac{j}{L})+\frac{\tilde{d}_{22}(\tilde{d}_{22}-\tilde{d}_{11})}{\tilde{c}^2}(\frac{i}{L})
\end{equation}
for $\tilde{c}=c_{11}=c_{22}$. In (\ref{tpcf1}) $d_{ij}$'s are
defined as $D_{ij}=(C-E)_{ij}\equiv d_{ij}$ and in (\ref{tpcf2})
$\tilde{d}_{ij}$'s are the same elements for the case
$c_{11}=c_{22}$.

\section{The Yang-Lee zeros of the canonical partition function}

In this section we will focus on some of the steady state
properties of our branching-coalescing model first studied in
\cite{farhad2}. We will first calculate the canonical partition
function of our model with the following non-zero rates:
\begin{equation}
\label{P1}
\begin{array}{c}
w_{32}=q,w_{23}=\frac{1}{q},w_{34}=q,w_{43}=\Delta
q,w_{24}=\frac{1}{q},w_{42}=\frac{\Delta}{q},\alpha,\gamma.
\end{array}
\end{equation}
It can easily be verified that this model is an special case of
the BCRW. In this model we have $\beta =0$ and the constraints
(\ref{Rules}) also give $\rho=\frac{\Delta}{1+\Delta}$ and
$\alpha=(\frac{1}{q}-q+\gamma)\Delta$. By introducing the
canonical ensemble we restrict ourself to the case where the total
number of particles on the chain $M$ is fixed. As we mentioned, it
has been shown that the classical Yang-Lee theory of the
equilibrium phase transitions can successfully be applied to the
driven diffusive systems to predict their non-equilibrium phase
transitions. By finding the roots of the canonical partition
function of our model, we will investigate its phase transitions.
In order to calculate the canonical partition function we will use
the quadratic algebra of the model
\begin{equation}
\begin{array}{l}
\label{FinalBulkAlgebra}
[C,\bar{C}] = [E,\bar{E}] = 0 \\ \\
\bar{E}C-E\bar{C} =(q+q \Delta+ q^{-1}) EC - q(1+\Delta)  E^{2} -
q^{-1} C^{2} \\ \\
\bar{C}E-C\bar{E} =(q^{-1}+q^{-1} \Delta+ q)  CE -q^{-1}(1+\Delta)
E^{2} - q C^{2}\\ \\ \langle W
\vert((\alpha+\gamma)E+\bar{E}-\gamma C)=\langle W \vert \bar
{C}=0 \; , \; \bar{E} \vert V \rangle =\bar{C}\vert V \rangle=0
\end{array}
\end{equation}
and its representation for $q^2\neq 1+\Delta$
\begin{equation}
\begin{array}{c}
\label{RepBulk}
C=\left(\begin{array}{cc}
1+\Delta & 0  \\
0 & q^{2}
\end{array} \right) \; , \;
E=\left(\begin{array}{cc}
1 & \lambda  \\
0 & q^{2}
\end{array} \right),
\bar{E}=\left(\begin{array}{cc}
\frac{q^{2}-1}{q}\Delta & -\frac{\Delta \lambda}{q} \\
0 & 0 \end{array} \right)\\ \\
\bar{C}=0 \; , \; \vert V \rangle=\left(\begin{array}{c}
\frac{\lambda}{q^{2}-1}\\1
\end{array} \right),\langle W \vert=\Bigl(\frac{-q^2\alpha\lambda^{-1}}
{\gamma(1+\Delta)-q\Delta},1 \Bigr).
\end{array}
\end{equation}
For convenience we will use our new notation in which $\gamma$
(instead of $\beta$ in \cite{farhad2}) is the extraction rate of
particles from the left boundary. It can easily be seen from
(\ref{RepBulk}) that the operators $C$ and $E$ have the following
properties
\begin{equation}
\label{RepProp} E(C-E)=(C-E) \ \ , \ \ (C-E)^i=\Delta^{i-1}(C-E)
\end{equation}
We define the canonical partition function $Z_{L,M}$ as the sum
over the steady states weights with fixed number of particles $M$.
Using (\ref{FinalBulkAlgebra})-(\ref{RepProp}) and the approach
used in \cite{lpk,farhad1} we obtain
\begin{equation}
\label{CPF2}
Z_{L,M}=\frac{q^2\alpha(1-\frac{\Delta}{q^2-1})\Delta^M}{\gamma(1+\Delta)-q\Delta}
\sum_{i=0}^{L-M}q^{2i}C^{M-1}_{L-i-1}
\end{equation}
where $C^{j}_{i}=\frac{i!}{j!(j-i)!}$ is the binomial coefficient.
Let us now examine the zeros of the canonical partition function
(\ref{CPF2}) as a function of $q$ for fixed $L$, $M$, $\alpha$,
$\gamma$ and $\Delta$. Apart from the zeros generated by the
factor behind the sum in (\ref{CPF2}), the numerical estimates of
the roots in the complex-$q$ plane are plotted in Figure (1) for
$L=500$ and $M=300$ (grey dots).
\begin{figure}[htbp]
\setlength{\unitlength}{1mm}
\begin{picture}(0,0)
\put(-5,30){\makebox{$\scriptstyle Im(q)$}}
\put(56,-2){\makebox{$\scriptstyle Re(q)$}}
\end{picture}
\centering
\includegraphics[height=5cm] {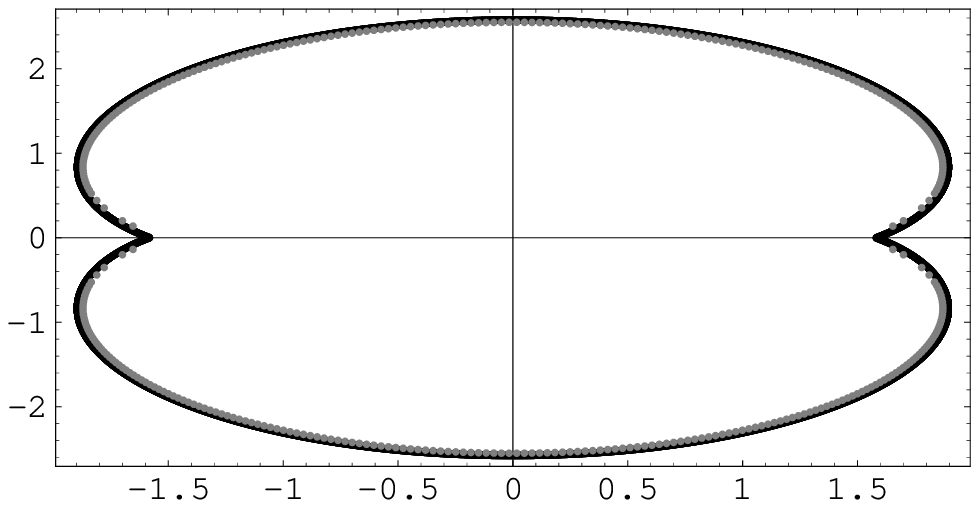}
\caption{Numerical estimates (grey dots) for the roots of
$\sum_{i=0}^{L-M}q^{2i}C^{M-1}_{L-i-1}$ besides the line of zeros
(dark line) given by (\ref{LOZ}) for $L=500$ and $M=300$ in the
complex-q plane. The exact critical point is
$q_c=\frac{1}{\sqrt{1-\rho}}=1.581$.}
\end{figure}
As can be seen from Figure (1) the roots accumulate towards a
critical point on the positive real-$q$ axis which will be
calculated shortly. In order to calculate the line of zeros
analytically we need the thermodynamic behavior of the canonical
partition function (\ref{CPF2}). This can be done using the
steepest decent method. We find
\begin{eqnarray}
\label{zfunc} Z_{L,M}\sim\cases{
\frac{q^{2L}\Delta^M}{(q^{2L}-1)^M},&$\rho<1-q^{-2}$\cr \frac{\rho
C_{L}^{M}\Delta^M}{1+q^2(\rho-1)},&$\rho>1-q^{-2}$\cr}
\end{eqnarray}
in which $\rho=\frac{M}{L}$. The critical point is obviously
$q_c=\frac{1}{\sqrt{1-\rho}}$. Defining the extensive part of the
free energy as
\begin{equation}
\label{gfunc}
g=\lim_{L,M \rightarrow \infty}\frac{1}{L}\ln
Z_{L,M}
\end{equation}
one can calculate the line of zeros from
\begin{equation}
\label{loz}
Re \; g_1=Re \; g_2
\end{equation}
in which $g_1$ and $g_2$ are the free energy functions on the left
and the right-hand side of the critical point \cite{gross}. After
some calculations using (\ref{zfunc})-(\ref{loz}) we obtain the
following equation for the line of zeros in the thermodynamic
limit:
\begin{equation}
\label{LOZ}
\frac{x^2+y^2}{[(x^2-y^2-1)^2+(2xy)^2]^{\frac{\rho}{2}}}=
\frac{(1-\rho)^{\rho-1}}{\rho^\rho}
\end{equation}
in which we have defined $x:=Re(q)$ and $y:=Im(q)$. This function
is plotted in Figure (1) for $L=500$, $M=300$ and
$\rho=\frac{M}{L}=0.6$ (dark line). It is seen that the line of
zeros lays on the numerical estimates of the zeros. The function
(\ref{LOZ}) intersects the positive real-q axis at the critical
point $x_c=\frac{1}{\sqrt{1-\rho}}$. We can also predict the order
of transition by fining the density of roots near the critical
point. For $0<y \ll 1$ and $(x-\frac{1}{\sqrt{1-\rho}})\ll 1$ we
find that the line of zeros is actually a straight line $y\sim
x-\frac{1}{\sqrt{1-\rho}}$ with the slope $\frac{\pi}{4}$. The
density of zeros on this line can be computed from
\begin{equation}\mu(s)=\frac{1}{2
\pi}\vert \frac{\partial }{\partial s }Im(g_1-g_2)\vert
\end{equation}
where $s$ is the distance from the transition point \cite{gross}.
It turns out that the density of zeros as a function of the
distance in the vertical direction is proportional to $y$ so in
$y\rightarrow 0$ limit the density of zeros vanishes. This
expresses a
second-order phase transition at the critical point. \\
In what follows we will calculate the density profile of particles
on the chain in each phase using the canonical ensemble. In the
canonical ensemble, the density of particles at site $i$ can be
written as
\begin{equation}
\langle \rho_i \rangle= \frac{\sum_{\cal C}\tau_iP^*({\cal
C})}{\sum_{\cal C} P^*({\cal C})}.
\end{equation}
Note that the sums are over the configurations with fixed number
of particles $M$. By using (\ref{FinalBulkAlgebra})-(\ref{RepProp})
and after some calculations we obtain
\begin{equation}
\label{DP} \langle \rho_i
\rangle=\frac{\sum_{k=0}^{L-Max(i,M)}q^{2k}C_{L-k-2}^{M-2}+C_{i-2}^{M-1}q^{2(L-i)}
\theta(M<i)}{\sum_{k=0}^{L-M}q^{2k}C_{L-k-1}^{M-1}}
\end{equation}
in which $\theta(\cdots)$ is the ordinary Heaviside function. In
Figure (2) we have plotted (\ref{DP}) for $L=500$, $M=300$ and two
values of $q$. With this choice of $L$ and $M$ the critical point
is $q_c=\frac{1}{\sqrt{1-\rho}}=1.581$.
\begin{figure}[htbp]
\setlength{\unitlength}{1mm}
\begin{picture}(0,0)
\put(-3,27){\makebox{$\scriptstyle \langle \rho_i \rangle$}}
\put(45,-1){\makebox{$\scriptstyle i$}}
\end{picture}
\centering
\includegraphics[height=5cm] {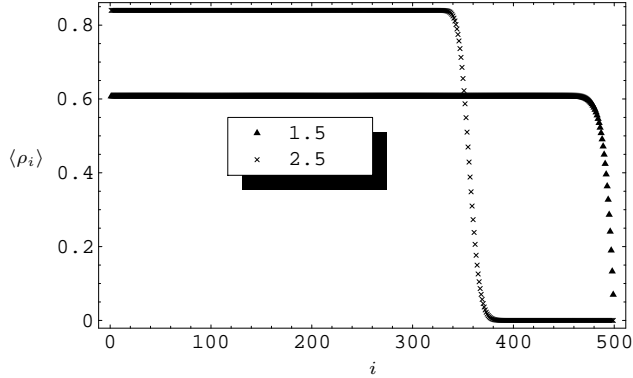}
\caption{Plot of the equation (\ref{DP}) for $L=500$, $M=300$ and
two values of $q$ as it is shown. Below the transition point
($q_c=1.581$) the density profile has an exponential behavior
while it is an error function above the critical point.}
\end{figure}
As can be seen for $q=1.5<q_c$ the density profile of particles is
constant through the bulk of the chain equal to
$\rho=\frac{M}{L}=0.6$ and drops to zeros only near the right
boundary. For $q=2.5>q_c$ the density profile of particles is a
shock. The high-density region ($\rho_{High}=1-q^{-2}=0.84$)
extending from $i=1$ to $i=\frac{\rho}{1-q^{-2}}L$ is separated by
a sharp interface from a low-density region ($\rho_{Low}=0$).
Using the steepest decent method one can easily show that in the
power-law phase ($q<q_c$) the density profile of particles
$\langle \rho_i \rangle$ has the following analytical form
\begin{equation}
\langle \rho_i \rangle=\rho(1-e^{-\frac{L-i+1}{\xi}})
\end{equation}
in which $\xi$ is the characteristic correlation length equal to
$\xi=\vert\ln(q^2(1-\rho))\vert^{-1}$.

\section{Generalization: Double shock structures}

In our non-conserving driven diffusive model \cite{farhad2} only
the left boundary was assumed to be open so that the particles
could enter and leave the system from the first site of the chain.
In this case the time evolution of the single shock measure was
equivalent to that of a random walker on a one-dimensional lattice
with reflecting boundaries. In this section we will introduce a
simple generalization of this model in which the particles are
allowed to enter and leave the system from both ends of the chain.
Our main objective will be to investigate the possibility of the
existence of double shock structures and the way that they evolve
in time. Defining a double shock measure, we will see that on
special manifold of parameters its time evolution is equivalent to
that of two random walkers on a one-dimensional lattice with
reflecting boundaries.\\ Let us assume that apart from the
interactions given by (\ref{P1}), the following reactions take
place at the right boundary
\begin{equation}
\label{P2}
\begin{array}{lcl}
\mbox{injection at the last site with rate}& \; \; &\delta\\
\mbox{extraction at the last site with rate}& \; \; &\beta.
\end{array}
\end{equation}
On a chain of length $L$ we define an uncorrelated double shock
measure as
\begin{equation}
\label{DSM} \vert m,n \rangle= \left( \ba{c} 1-\rho_1 \\
\rho_1 \ea \right)^{\otimes m} \otimes \left( \ba{c} 1-\rho_2 \\
\rho_2 \ea \right)^{\otimes n-m-1} \otimes \left( \ba{c} 1-\rho_1 \\
\rho_1 \ea \right)^{\otimes L-n+1}
\end{equation}
in which  $m+1 \leq n$, $0 \leq m \leq L$ and $1\leq n \leq L+1$.
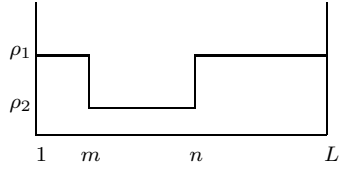
\begin{figure}[htbp]
\begin{center}
\begin{picture}(120,70)
\put(10,10){\line(0,1){50}} \put(10,10){\line(1,0){110}}
\put(30,20){\line(1,0){40}} \put(70,20){\line(0,1){20}}
\put(70,40){\line(1,0){50}} \put(120,10){\line(0,1){50}}
\put(10,40){\line(1,0){20}} \put(30,20){\line(0,1){20}}
\put(10,0){\footnotesize 1} \put(68,0){\footnotesize $n$}
\put(119,0){\footnotesize $L$} \put(0,40){\footnotesize $\rho_1$}
\put(0,20){\footnotesize $\rho_2$} \put(27,0){\footnotesize $m$}
\end{picture}
\caption[profig]{Density profile of a double shock measure}
\label{profig}
\end{center}
\end{figure}
We have sketched the density profile of a sample double shock
measure in Figure (3). It can easily be verified that the time
evolution of the double shock measure (\ref{DSM}) is equivalent to
that of two random walkers provided that
\begin{equation}
\label{DSMC} \rho_1=\frac{\Delta}{1+\Delta}\; ,\; \rho_2=0 \;, \;
\alpha=(q^{-1}-q+\gamma)\Delta \; , \;
\delta=(q-q^{-1}+\beta)\Delta.
\end{equation}
Under these conditions the time evolution equation of (\ref{DSM})
generated by the Hamiltonian of our generalized model will be
given by the following equations
\begin{equation}
\begin{array}{l}
\label{TEE} -H \vert m,n\rangle =q \vert m-1,n\rangle
+q^{-1}(1+\Delta)\vert m+1,n\rangle+q^{-1}\vert
m,n+1\rangle+\\q(1+\Delta)\vert
m,n-1\rangle-(2+\Delta)(q^{-1}+q)\vert m,n\rangle , \\
\mbox{for} \;\;\; m\neq 0,n\neq L+1,n\neq m+1 \\ \\
-H\vert 0,n\rangle=\alpha(\frac{1+\Delta}{\Delta})\vert 1,n\rangle
+q^{-1}\vert 0,n+1\rangle+q(1+\Delta)\vert 0,n-1\rangle-\\
(\alpha(\frac{1+\Delta}{\Delta})+q^{-1}+q(1+\Delta))\vert
0,n\rangle,\\ \mbox{for} \;\;\; n\neq 1,L+1\\ \\
-H\vert m,L+1\rangle=q^{-1}(1+\Delta)\vert m+1,L+1\rangle+q\vert
m-1,L+1\rangle+ \\ \delta(\frac{1+\Delta}{\Delta})\vert
m,L\rangle-(\delta(\frac{1+\Delta}{\Delta})+q+q^{-1}(1+\Delta))
\vert m,L+1\rangle,\\ \mbox{for} \;\;\; m \neq 0,L \\ \\
-H\vert 0,L+1\rangle=\alpha(\frac{1+\Delta}{\Delta})\vert
1,L+1\rangle+\delta(\frac{1+\Delta}{\Delta})\vert 0,L\rangle-\\
\frac{(\alpha+\delta)(1+\Delta)}{\Delta}\vert 0,L+1\rangle\\ \\
-H\vert m,n\rangle=0 \;\;\; \mbox{for} \;\;\; n=m+1.
\end{array}
\end{equation}
The last equation indicates that the model has a trivial steady
states in which the density of particles at all sites is equal to
$\rho_1=\frac{\Delta}{1+\Delta}$. This is associated with a
one-dimensional representation of the quadratic algebra of this
model. As can be seen from (\ref{TEE}) the shock fronts move like
two biased random walkers with different hopping rates. The shocks
are also reflected from the boundaries. In the most general form
it can easily be seen that for an open boundary model with the
following non-vanishing rates
\begin{equation}
\label{DSMR} w_{23},\; w_{32},\; w_{24},\; w_{42},\; w_{34},\;
w_{43} ,\; \alpha,\; \beta,\; \gamma,\; \delta
\end{equation}
the above-mentioned approach can be used to show that the double
shock structures of kind (\ref{DSM}) exist if the reaction rates
satisfy the following constraints:
\begin{equation}
\begin{array}{l}
\label{DSMCG}
\rho_1=\frac{w_{42}+w_{43}}{w_{24}+w_{34}+w_{42}+w_{43}} \; , \; \rho_2=0 \\ \\
w_{32}=\frac{w_{42}(w_{24}+w_{34})}{w_{42}+w_{43}}
\; , \; w_{23}=\frac{w_{43}(w_{24}+w_{34})}{w_{42}+w_{43}}\\ \\
\alpha=\frac{w_{42}+w_{43}}{w_{24}+w_{34}}(w_{24}-
\frac{2w_{42}(w_{24}+w_{34})}{w_{42}+w_{43}}+
\frac{(w_{24}+w_{34})(w_{34}+w_{42})}{w_{24}+w_{34}+w_{42}+w_{43}}+\gamma) \\ \\
\delta=\frac{w_{42}+w_{43}}{w_{24}+w_{34}}(-w_{24}+
\frac{2w_{42}(w_{24}+w_{34})}{w_{42}+w_{43}}-
\frac{(w_{24}+w_{34})(w_{34}+w_{42})}{w_{24}+w_{34}+w_{42}+w_{43}}+\beta).
\end{array}
\end{equation}
In this case the time evolution equations for $\vert m,n \rangle$
given by (\ref{DSM}) will be quite similar to (\ref{TEE}). One can
readily check that for our generalized model, defined by
(\ref{P1}) and (\ref{P2}), the conditions (\ref{DSMCG}) reduce to
(\ref{DSMC}). The steady state of the system can be obtained from
the superposition of double shock measures \cite{farhad4};
however, the system has a trivial steady state in which the
particles occupy the sites of the chain with the uniform
probability $\rho_1$ and as before this is associated with a
one-dimensional representation of its quadratic algebra.

\section{Concluding remarks}

Our calculations in this paper can be divided into three parts. In
the first part of the paper we have shown that using the MPF the
steady state of the ASEP, BCRW and AKGP can be expressed by
two-dimensional representations of their quadratic algebras
provided that some constraints hold on the reaction rates. These
constraints introduce specific manifold of parameters. On these
manifolds the stationary measures are given by superpositions of
Bernoulli shock measures. Having the explicit form of the
two-dimensional representations several physical quantities, such
as correlation functions, can easily be calculated.\\
The BCRW is a generalized form of the model that we had studied in
\cite{farhad2}. The second part of our calculations is devoted to
the detailed study of the phase transition and also the shock
formation in this model as an special case of the BCRW. We have
introduced a canonical ensemble in which the number of particles
on the chain is fixed. Using the properties of the quadratic
algebra the canonical partition function of the model is
calculated exactly. The thermodynamic behavior of this function
reveals a phase transition which depends on $q$ and $\rho$ i.e.
one of the reaction rates and the density of particles on the
chain. In order to study this phase transition we have used the
classical Yang-Lee theory. Our calculations show that the Yang-Lee
zeros of the canonical partition function as a function of the
complex variable $q$ lie on a curve which intersect the positive
real-q axis at an angle $\frac{\pi}{4}$ at the critical point
$q_c=\frac{1}{\sqrt{1-\rho}}$. According to the Yang-Lee theory of
phase transitions, this is reminiscent of a second-order phase
transition. The nature of each phase can be studied by calculating
the density profile of the particles in the steady state. For $q <
q_c$ the density of particles is constant everywhere on the chain
except near the right boundary where it drops to zero
exponentially with the correlation length $\xi=\vert
q^2(1-\rho)\vert^{-1}$ while it has a shock structure for $q > q_c$.\\
In the third part of our calculations we have shown that double
shock structures can evolve in a general branching-coalescing
model with open boundaries. In this case the boundary parameters
lie on specific manifolds of parameters determined by
(\ref{DSMCG}). A detailed study of the double shock properties
will be given elsewhere \cite{farhad4}.\\
The exact solution of the BCRW without any constraints on its
reaction rates still remains an open question. The exact steady
state of the model defined by the non-vanishing rates (\ref{DSMR})
under the conditions (\ref{DSMCG}) from the MPF point of view is
under consideration \cite{farhad4}.

\section*{Appendix}

It turns out that the representation of the quadratic algebra
of the BCRW in the case $c_{11}\neq c_{22}$ (see (\ref{Rep1})) is
$$
\begin{array}{c}
C=\left(\begin{array}{cc}
\frac{w_{23}}{1-\rho} & 0  \\
0 & (1-\rho)w_{32}+\rho w_{34}
\end{array} \right) \; , \;
E=\left(\begin{array}{cc}
w_{23} & \lambda  \\
0 & (1-\rho)w_{32}+\rho w_{34}
\end{array} \right),\\ \\
\bar{E}=\left(\begin{array}{cc}
-w_{43}(w_{23}-w_{32}(1-\rho)-w_{34}\rho) & -\lambda\frac{w_{23}\rho}{1-\rho} \\
0 & 0 \end{array} \right) \; , \; \bar{C}=0 \\ \\
\vert V \rangle=\left(\begin{array}{c}
\frac{\lambda(1+\frac{\beta}{w_{43}})}{\beta+w_{32}(1-\rho)+w_{34}\rho-w_{23}}\\1
\end{array} \right),\langle W \vert=\Bigl(\frac{\alpha\rho(w_{32}(\rho-1)-w_{34}\rho)}
{\lambda(\alpha+w_{23}\frac{\rho}{\rho-1}+\rho(w_{32}(1-\rho)+w_{34}\rho))},1 \Bigr).
\end{array}
$$
and in the case $c_{11}=c_{22}$ (see (\ref{Rep2})) it is
$$
\begin{array}{c}
C=\left(\begin{array}{cc}
w_{32}(1-\rho)+w_{34}\rho & \lambda \\
0 & w_{32}(1-\rho)+w_{34}\rho
\end{array} \right) \\ \\
E=\left(\begin{array}{cc}
(1-\rho)(w_{32}(1-\rho)+w_{34}\rho) & \frac{-\eta}{\rho(w_{32}(1-\rho)+w_{34}\rho)}  \\
0 & w_{32}(1-\rho_1)+w_{34}\rho
\end{array} \right), \\ \\
\bar{E}=\left(\begin{array}{cc}
\rho^2(w_{32}(1-\rho)+w_{34}\rho)^2 &\eta \\
0 & 0 \end{array} \right) \; , \;
\bar{C}=0 \; , \; \vert V \rangle=\left(\begin{array}{c}
v_1\\1
\end{array} \right),\langle W \vert=\Bigl(w_1,1 \Bigr).
\end{array}
$$
in which
$$
\begin{array}{l}
v_1=-\frac{\eta\beta+w_{32}(\eta+\lambda
\beta)\rho-(w_{32}-w_{34})(\eta+\lambda\beta)\rho^2}
{\rho^2(w_{32}(1-\rho)+w_{34}\rho)^2(\beta+\rho(w_{32}(1-\rho)+w_{34}\rho))} \\
w_1=\frac{\alpha\rho^2(w_{32}(1-\rho)+w_{34}\rho)^2}
{\eta\alpha+\lambda\rho(w_{32}(1-\rho)+w_{34}\rho)(\alpha(1-\rho)+\rho^2(w_{32}(1-\rho)+w_{34}\rho))}
\end{array}
$$
and $\lambda$ and $\eta$ are arbitrary parameters.
The representation of the quadratic algebra for the AKGP is also found to be
$$
\begin{array}{c}
C=\left(\begin{array}{cc}
w_{43} & 0  \\
0 & w_{13}
\end{array} \right) \; , \;
E=\left(\begin{array}{cc}
0 & \lambda  \\
0 & w_{13}
\end{array} \right),
\bar{E}=\left(\begin{array}{cc}
0 & \lambda(w_{13}-w_{43}) \\
0 & 0 \end{array} \right) \; , \; \bar{C}=0 \\ \\
\vert V \rangle=\left(\begin{array}{c}
\frac{\lambda(w_{43}-w_{13}+\beta)}{\beta w_{43}}\\1
\end{array} \right),\langle W \vert=\Bigl(\frac{-\alpha w_{13}}{\lambda(w_{13}-w_{43}+\alpha)},1 \Bigr).
\end{array}
$$
and
$$
\begin{array}{c}
C=\left(\begin{array}{cc}
w_{13} & \lambda  \\
0 & w_{13}
\end{array} \right) \; , \;
E=\left(\begin{array}{cc}
0 & \lambda  \\
0 & w_{13}
\end{array} \right),
\bar{E}=\left(\begin{array}{cc}
0 & -\lambda w_{13} \\
0 & 0 \end{array} \right) \; , \; \bar{C}=0 \\ \\
\vert V \rangle=\left(\begin{array}{c}
\frac{\lambda}{\beta}\\1
\end{array} \right),
\langle W \vert=\Bigl(\frac{\alpha w_{13}}{\lambda(w_{13}-\alpha)},1 \Bigr).
\end{array}
$$
for the case $c_{11}\neq c_{22}$ and $c_{11}=c_{22}$ respectively.

\end{document}